\documentstyle[12pt]{article}

         \def\la{\label}
         \def\rf{\ref}

         \def\be{\begin{equation}}
         \def\bea{\begin{eqnarray}}
         \def\ee{\end{equation}}
         \def\eea{\end{eqnarray}}
         \def\se{\section}
         
         \def\o{\over}
         \def\a{\alpha}
         \def\b{\beta}
\begin{document}
\begin{titlepage}
\vspace*{20mm}
\begin{center} {\Large \bf A Remark on Integrability of Stochastic Systems
\\ \vskip 0.35cm
Solvable by Matrix Product Ansatz}\\
\vskip 1cm
\centerline {\bf V. Karimipour
\footnote {e-mail:vahid@netware2.ipm.ac.ir}}
\vskip 1cm
{\it Department of Physics, Sharif University of Technology, }\\
{\it P.O.Box 11365-9161, Tehran, Iran }\\
{\it Institute for Studies in Theoretical Physics and Mathematics,}\\
{\it P.O.Box 19395-5746, Tehran, Iran}\\
\end{center}

\vskip 1.5cm
\begin{abstract}
\noindent
Within the Matrix Product Formalism we have already introduced a multi-species
exclusion process \cite {tra},\cite {ring}, in which different particles
hop with different rates and fast particles stochastically overtake slow ones.
In this letter we show that
on an open chain, the master equation of this process can be exactly solved
via the coordinate Bethe ansatz. It is shown that the N-body S-matrix
of this process is factorized into a product of two-body S-matrices, which
in turn satisfy the quantum Yang-Baxter equation (QYBE). This solution is
to our knowledge, a new solution of QYBE.
\end{abstract}
\end{titlepage}
\newpage
\section{Introduction}
The static Matrix Product Ansatz (MPA) as formulated in \cite{KS}, relates the
problem
of finding the steady state of a one dimensional homogenous stochastic system
, to a quadratic algebra and its representations.\\
Looking back at the history of the subject, one can look at static MPA and its
final formulation by Krebs and Sandow\cite{KS}, as an abstractization of some
recursion relations between certain quantities for systems of different sizes.
In this way the technique of using recursion relations \cite{aa}\cite{bb}
acquires a purely algebraic character, and the results obtained by the former
method \cite{aa}\cite{bb} are rederived \cite{dehp} by computing matrix
elements
of product of operators. Conceptually MPA is the generalization of the ordinary
Bernoulli measure where matrix elements of operators replace
numbers to produce correlations.\\
Many of
the questions regarding the steady state and the correlation functions can then
be answered at least in principle, by purely algebraic manipulations.
We say "in principle", since in general, finding the representations of the
algebra may be as difficult as the original problem. The special cases where
the static MPA has been used for calculating concrete physical quantities
include the one-species asymmetric simple exclusion process (ASEP) on a
ring and on an open chain with and without an impurity
\cite{dehp},\cite{djls},,\cite{m},\cite{lpk},\cite{de}(see also \cite{dr},\cite{evans} and  references therein).
The works on the multi-species asep include \cite{ritt} \cite{e},\cite{ring},
and \cite{tra}. When one
comes to the dynamical MPA \cite{ss} \cite{s}, the corresponding algebra
becomes time-dependent and
much more involved.
Even when all the particles are identical,
only in certain limited cases the dynamical algebra has been used for
calculating
physical quantities, namely the symmetric exclusion process \cite{ss} and a
non-equilibrium model of spin relaxation \cite{s}. In these two cases the
algebra is
simple enough to be used for finding the spectrum of the corresponding quantum
Hamiltonian and $r$-point density correlators.
These two cases happen to be integrable. In fact they belong to a 10-parameter
family of diffusion-reaction systems which are equivalent through a
similarity transformation to a generalized Heisenberg chain \cite{s}, which is
known to be integrable. \\
For a general stochastic process, dynamical MPA, like the
static one is just a rewriting
of the master equation and its real power is revealed only when one can solve
the resulting algebraic relations or can find a representation of the
algebra.\\
It is useful to agree upon a notion, however vague, of MPA solvability. Let us
call a process static-MPA (dynamic-MPA) solvable, if a nontrivial
representation of the static (dynamic) MPA algebra can be found, or the
ensuing recursion relations can be solved.\\
At present there is no definite connection between static or
dynamic MPA-solvability and integrability, although from the above examples it
appears
that the dynamic-MPA solvable systems are integrable, since for these models
the entire spectrum has been found. However it is not known what kind of
integrable systems are MPA solvable.\\
To see if there is any  connection between MPA-solvability and
integrability, it is desirable to study a number of examples which hint to
such a connection and to investigate the common properties of these examples.\\
In this letter we provide an example of
a multi-species stochastic process which on the one hand is related to a very
simple algebra in the MPA
formalism and on the other hand is integrable in the sense that its N-particle
S-matrix
is factorized into a product of two-body S-matrices, which
in turn satisfy the quantum Yang-Baxter equation (QYBE).
The algebra is generated by $p+1$ generators $ E $ and $ D_i, i = 1\cdots p $
with relations \cite{tra}  \begin{eqnarray}\la{B} D_i E &=& {1\over v_i} D_i +
E  \hskip 1cm i=1...p\\
 D_j D_i &=& {v_i D_j - v_j D_i \over {v_i - v_j }}\hskip 1cm j>i\end{eqnarray}
where $ v_1\leq v_2 \leq \cdots \leq v_p $ are finite non-zero real numbers.
In this process a particle of type $ j $
hops with rate $ v_j $ and when it encounters a particle of type $i $,
with $ v_i < v_j $
the two particles interchange their sites, with a rate $ v_j - v_i $ as if
the fast particle stochastically overtakes the slow one.
This model seems to be natural as a simple
model of one-way traffic flow.
We have shown elsewhere \cite{tra},\cite{ring} that on a finite chain this
process is MPA-solvable in the above mentioned sense. We now consider the
same process on an open chain and show that the master equation of this process
can be solved via the coordiante Bethe ansatz. That is, by a suitable
transformation
we turn the master equation into an eigenvalue equation, and find the
eigenvalues and eigenfunctions of this equation, the latter being in the
form of Bethe wavefunctions. In this case, since we have $p$ species of
particles,
the two particle S-matrix is not a c-number as in \cite{Gs},\cite{Sch} but is a
$ p^2\times p^2$ matrix. Thus the factorizability of the N-particle S-matrix
puts a stringent condition on this two-body S-matrix, in the form of
the Yang-Baxter
equation. We will show that this is indeed the case, and as a byproduct
provide a new solution of QYBE with spectral parameter.
\se{The Master Equation}
Suppose there are $N$ particles of different types on an infinite one
dimensional
chain. The basic objects that we are interested in, are denoted by
$ P_{\a_1,\a_2,\cdots \a_N}(x_1, x_2, \cdots x_N)$ which are the
probabilities of finding at time $t$, a particle
of type $ \alpha_1 $ at site $ x_1 $, a particle of type
$ \alpha_2 $ at site $ x_2 $,
etc. Following \cite{Sch}, these functions to define probabilities
only in the {\it physical region} $ x_1 < x_2 < \cdots < x_N $.
The hypersurfaces where any of the two adjacent coordinates are equal are the
boundaries of the physical region. In the subset of the physical region where
$ x_{i+1}- x_i > 1, \forall i $ we have free hopping of particles and the
master equation is written
as
\bea\la{1}
{\partial\over \partial t}P_{\a_1,\a_2,\cdots \a_N}(x_1, x_2,\cdots ,x_N;t)
&=& v_{\a_1} P_{\a_1,\a_2,\cdots \a_N}(x_1-1, x_2,
\cdots ,x_N;t) +\cdots\cr &+& v_{\a_N}P_{\a_1,\a_2,\cdots \a_N}
(x_1, x_2,\cdots ,x_N-1;t)\cr
&-& (v_{\a_1}+\cdots v_{\a_N}) P_{\a_1,\a_2,\cdots \a_N}
(x_1, x_2,\cdots , x_N;t ).\eea
We assume hereafter for convenience that the time variable has been rescaled
so that all the hopping rates are dimensionless quantities, i.e. they are
ratios of actual hopping rates to some standard hopping rate.\\
Extending the above equation to the whole physical region, where
$ x_{i+1} - x_{i} \geq 1 $, causes
terms on the boundary surfaces to appear on the right hand side of the master
eqution, which should be fixed by a choice for the boundary condition.
The choice
of boundary condition in fact determines the type of interaction between
particles (see \cite{AKK}). On any such surface where for example $ x_{i+1}
 - x_{i} = 1 $ for some $ i $, we supplement the
above equation with the following boundary condition, where we suppress for
simplicity the time variable and all the other coordinates: \ \ \
\be\la{2}
v_{\b}P_{\a, \b}(x,x)=v_{\a} P_{\a, \b}(x, x+1)+ (v_{\b}-v_{\a})
P_{\b, \a}(x,x+1) \hskip 1cm \b\geq \a\ee \be\la{3}
P_{\b, \a}(x,x)= P_{\b, \a}(x, x+1)\hskip 1cm \b\geq \a\ee
The master equation (\rf{1}) and the boundary conditions (\rf{2},\rf{3}),
replace
the very large number of equations which one should write by considering the
multitude of cases which arise according to which group of particles
are adjacent to each other. Instead of giving a general proof we
provide a few examples in the two and three particle sectors. \\
Consider the two particle sector. From the definition of the process we have:
\be\la{4}
{\partial\over \partial t}P_{12}(x,x+1)=v_{1} P_{12}(x-1, x+1)+ (v_{2}-v_{1})
P_{21}(x,x+1)- v_{2} P_{12}(x,x+1)\ee
\be\la{5}
{\partial\over \partial t}P_{21}(x,x+1)=v_{2} P_{21}(x-1, x+1)-
(v_{1}+(v_2-v_1)) P_{21}(x,x+1)\ee
These are exactly the equations which are obtained from combination of (\rf{1})
and (\rf{2}, \rf{3}).
As a couple of examples in the three particle sector, consider
$ P_{123}(x, x+1, x+2) $ and $ P_{213}(x, x+1, x+2) $. From the definition
of the process we have
\bea \la{8}
{\partial\over \partial t}P_{123}(x,x+1,x+2)&=&v_{1} P_{123}(x-1, x+1, x+2)
+ (v_2 - v_1) P_{213}(x,x+1,x+2) \cr &+& (v_3 - v_2) P_{132}(x,x+1,x+2)-
v_3 P_{123}(x,x+1,x+2)\eea
\bea\la{9}
{\partial\over \partial t}P_{213}(x,x+1,x+2)&=&v_{2} P_{213}(x-1, x+1, x+2)
+ (v_3 - v_1) P_{231}(x,x+1,x+2) \cr &-& (v_3 + (v_2-v_1)) P_{213}(x,x+1,x+2)\eea
These are precisely obtained from combination of (\rf{1}) and (\rf{2}, \rf{3}).
This pattern repeats in all sectors.
\se{ The Bethe Ansatz Solution}
We rewrite $P_{\a_1,\a_2,\cdots \a_N}(x_1, x_2,\cdots ,x_N;t)$ as
\be \la{10}P_{\a_1,\cdots \a_N}(x_1, \cdots ,x_N;t) = e^{-(E+v_{\a_1}+
\cdots v_{\a_N})t} (v_{\a_1})^{x_1}\cdots (v_{\a_N})^{x_N}\Psi_{\a_1,\cdots
\a_N}(x_1,\cdots ,x_N;t)\ee
and insert it into (\rf{1}) to obtain
\bea \la{11} &&\Psi_{\a_1,\a_2,\cdots \a_N}(x_1-1, x_2,\cdots ,x_N;t)+
\Psi_{\a_1,\a_2,\cdots \a_N}(x_1, x_2-1,\cdots ,x_N;t)\cr &+&\cdots \Psi_{\a_1,
\a_2,\cdots \a_N}(x_1, x_2,\cdots ,x_N-1;t) = -E \Psi_{\a_1,\a_2,\cdots \a_N}
(x_1, x_2,\cdots ,x_N;t)\eea
In terms of $\Psi$ the boundary conditions (\rf{2}, \rf{3}) are rewritten as
\be\la{12}
\Psi_{\a, \b}(x,x)=v_{\a} \Psi_{\a, \b}(x, x+1)+ {v_{\a}\o v_{\b}}
(v_{\b}-v_{\a}) \Psi_{\b, \a}(x,x+1)\hskip 1cm \b \geq \a\ee
\be\la{13}
\Psi_{\b, \a}(x,x)= v_{\a}\Psi_{\b, \a}(x, x+1)\hskip 1cm \b \geq \a\ee
Hereafter we use a compact notation as follows. ${\bf \Psi}$ is a
tensor of rank
N, whose components are $\Psi_{\a_1,\a_2,\cdots \a_N}(x_1, x_2,\cdots ,x_N)$.
The boundary conditions are written in the
form
\be \la{14} {\bf \Psi}(\cdots, \xi,\xi,\cdots) = {\bf b}_{k,k+1}{\bf
\Psi}(\cdots, \xi,\xi+1, \cdots)\ee
where ${\bf b}_{k,k+1}$ is the embedding of ${\bf b}$ (the matrix derived
from (\rf{12}, \rf{13}))
in the locations $k$ and $k+1$.
\be \la{15} {\bf b}_{k,k+1} = 1\otimes \cdots \otimes \underbrace {\bf b}
_{k,k+1} \otimes \cdots \otimes 1 \ee
In the two species case $ {\bf b} $ is equal to
\be \la{16} {\bf b} = \left( \begin{array}{cccc}v_1&.&.&.\\.&v_1&
{v_1\o v_2}(v_2-v_1)&. \\.&.&v_1&.\\.&.&.&v_2\end{array}\right)\ee
In the $p$-species case it is :
\be \la{bb} {\bf b}= \sum_{i,j}{\tilde v}_{ij} E_{ii}\otimes E_{jj} +
\sum_{i<j}{v_i\over v_j}(v_i-v_j) E_{ij}\otimes E_{ji}\ee
where
$ (E_{ij})_{k,l} = \delta_{ik} \delta_{jl} $ and ${\tilde v}_{ij} = v_{min
(i,j)}$.\\ To solve the eigenvalue equation (\rf{11}), we write $\Psi$ as a
Bethe wave function
\be\la{17}
{\bf \Psi} (x_1,\cdots ,x_N)=\sum_\sigma A_\sigma e^{i\sigma( {\bf p})\cdot
{\bf x}}.
\ee
Here {\bf x} and {\bf p} denote $n$-tuples of coordinates and momenta
respectively, and the summation runs over all the elements ${\sigma}$ of
the permutation group.
For each element $\sigma$ of the permutation group, the corresponding
coefficient (a tensor of rank N) is denoted by $ A_{\sigma}$.
Inserting (\rf{17}) in (\rf {11}), yields
\be\la{18}
\sum_\sigma A_\sigma e^{i\sigma ({\bf p})\cdot {\bf x}}\big(
\sum_j e^{-i\sigma (p_j)}+ E\big) = 0
\ee
From this, one obtains the eigenvalues (as one can remove $\sigma$
from the summations in the parenthesis)
\be\la{19}
E=-\big (e^{-ip_1}+ e^{-ip_2}+\cdots e^{-ip_N}\big).\ee
The Bethe wavefunction should also satisfy the boundary condition (\rf{14}).
Inserting (\rf{17}) in (\rf{14}) we obtain
\be \la{20}
\sum_{\sigma} e^{i\sum_{j\ne k, k+1}\sigma(p_j)x_j + i \big(\sigma(p_k)
+ \sigma(p_{k+1})\big)\xi}
\Bigg( 1 - e^{i\sigma(p_{k+1})}{\bf b}_{k,k+1}\Bigg) A_{\sigma} = 0 \ee
We now note that $ x_j $ and $ \xi$ are arbitrary and the coefficient of
$ \xi$ is symmetric with respect to the interchange of $ p_k $ and
$ p_{k+1}$. This interchange is effected by the element
$ \sigma_{k} $ of the permutation group. Thus symmetrizing
with respect to this interchange, we obtain
\be \la{21}
\Bigg( 1 - e^{i\sigma(p_{k+1})}{\bf b}_{k,k+1}\Bigg) A_{\sigma} +
\Bigg( 1 - e^{i\sigma(p_{k})}{\bf b}_{k,k+1}\Bigg) A_{{\sigma}\sigma_k} = 0 \ee
where ${\sigma}\sigma_k$ is the product of the elements $\sigma$ and $
\sigma_k$ in the permutation group and $ A_{{\sigma}\sigma_k} $ is the
corresponding coefficient.
Thus we obtain
\be \la{22}
A_{\sigma\sigma_k} = S_{k,k+1}(\sigma(p_k),\sigma(p_{k+1}))A_{\sigma}\ee
where
\be \la{23} {\bf S}_{k,k+1} (z_1, z_2)= 1\otimes \cdots \otimes \underbrace
{S(z_1,z_2)}_{k,k+1} \otimes \cdots \otimes 1 \ee
and
\be \la{24}
{\bf S}_(z_1, z_2)= - ( 1 - z_1{\bf b})^{-1} (1 - z_2 {\bf b})\ee
Here we have denoted $e^{ip_k}$ by $ z_k$.
$ S(z_1, z_2)$ is the two particle S-matrix. The above equation
allows all the $A_{\sigma}$'s
to be calculated recursively in terms of $ A_1$. The first few members of the
permutation group are $ 1, \sigma_1, \sigma_2, \sigma_1\sigma_2,
\sigma_2\sigma_1 $ and $ \sigma_1 \sigma_2\sigma_1= \sigma_2\sigma_1 \sigma_2 $.
Accordingly we find from (\rf{22})
\be \la{26}
A_{\sigma_1} = S_{12}(p_1, p_2)A_1 \hskip 1cm
A_{\sigma_2} = S_{23}(p_2, p_3)A_1 \hskip 1cm
A_{\sigma_1\sigma_2} = S_{23}(p_1, p_3)A_{\sigma_1}\ee
\be \la{27}
A_{\sigma_2\sigma_1} = S_{12}(p_1, p_3)A_{\sigma_2}\hskip 1cm
A_{\sigma_1\sigma_2 \sigma_1} = S_{12}(p_2, p_3)
A_{\sigma_1\sigma_2}. \ee
Moreover we have
\be \la{28}
A_{{\sigma_2}{\sigma_1}{\sigma_2}} = S_{23}(p_1, p_2)A_{\sigma_2\sigma_1}\ee
At this point one may think that the master equation (\rf{1}) can be solved
by Bethe ansatz for any type of matrix $ {\bf b} $ and hence for
any type of nearest
neighbour interaction. The answer is that we are not yet done with the Bethe
ansatz solution, since for consistency of the solution we should check that
different ways of obtaining a coefficient yeild the same result. It is
sufficient to check this consistensy only for two cubic elements of the
permutation group \cite{Ch}, namely $ \sigma_1\sigma_2\sigma_1
$ and $\sigma_2\sigma_1\sigma_2$ which are equal as
elements of this group. Thus we should have
\be \la{29}
A_{\sigma_1\sigma_2\sigma_1} = A_{\sigma_2\sigma_1\sigma_2}\ee
This yields via (\rf{26}) - (\rf{28}) the following Yang-Baxter
equation: \be \la{30}
S_{12}(p_2,p_3)S_{23}(p_1,p_3)S_{12}(p_1,p_2) = S_{23}(p_1,p_2)
S_{12}(p_1,p_3)S_{23}(p_2,p_3)\ee
which is crutial for consistency of the method.\\
{\bf Remark:} With the transformation $ R:=PS$, where $ P $
is the permutation operator, we obtain the more familiar form:
\be \la{C}
R_{23}(p_2,p_3)R_{13}(p_1,p_3)R_{12}(p_1,p_2) =
R_{12}(p_1,p_2)R_{13}(p_1,p_3)R_{23}(p_2,p_3)\ee
We can calculate the S-matrix from (\rf{17}) and (\rf{24}).
For example in the two-species
case  we find
\be \la{31}
S(z,w) = \left( \begin{array}{cccc}{1-v_1w\o v_1z-1}&.&.&.\\.&
{1-v_1w\o v_1z-1}&{v_1\o v_2}(v_2-v_1){w-z\o (1-v_1z)^2}&. \\.&.&
{1-v_1w\o v_1z-1}&.\\.&.&.&{1-v_2w\o v_2z-1}\end{array}\right)\ee
Note that the last diagonal element is different from the others.
We have checked that this matrix satisfies exactly equation (\rf{30}), although
its proof requires lenghty calculations which we will not reproduce
here. It has
moreover the following properties\\
{\bf 1}- In the special case where $ v_1 = v_2$  we have $ S(z,w) = {1-w\over
z-1}{\bf 1} $
which is nothing but the S-matrix found in \cite{Gs} for the single species
case. Note that in the latter case the S-matrix is a c-number and (\rf{30})
is satisfied trivially.\\
{\bf 2}- $S^{-1}(z,w) = S(w,z)$\\
{\bf 3}- $S(z,z) = {\bf 1} \ \ \ \ \forall\ \ \  v_1, v_2 $\\
One can easily obtain from (\rf {17}) and (\rf {24}), the S-matrix
for higher values of $ p$.
The final result is
\be \la{32}
S(z,w) = \sum_{i,j}^p {1-{\tilde v}_{ij} w\over {\tilde v}_{ij}z-1}
E_{ii}\otimes
E_{jj} + \sum _{i<j} {v_i\o v_j}(v_j-v_i){w-z\o (1-v_iz)^2} E_{ij}\otimes
E_{ji} \ee where $ {\tilde v}_{ij} = v_{min (i,j)}$.\\
In conclusion we have provided an example of a multi-species stochastic process
(the number of species being arbitrary), which is both static-MPA solvable
(see
the introduction) and integrable on an infinite chain. Integrability of the
model is ensured by testing a very stringent condition on the S-matrix, namely
the 2-particle S-matrix satisfies the Quantum Yang-Baxter equation. This
solution is to our knowledge a new solution of QYBE.
Given these facts, it would be quite interesting to see if this process is also
dynamic-MPA solvable. \\
At a concrete level and for sectors of low number of particles, one can use
the Bethe wave functions to calculate the same quantities that were calculated
in \cite {Sch}, namely the diffusion and drift constant, and see what is the
effect of overtaking on these characteristics. Questions specific to
the multispecies case
can also be answered. For example, given an initial situation of
particle 2, $l$ steps behind particle 1, what is the probability that at
a given time particle 2 overtakes particle 1, or what is the average passing
time.

\section{Aknowledgement}
I would like to thank the referees for their constructive comments. I also
thank G. M. Sch{\"u}tz for his valuable comments through an E-mail
correspondence.

\end{document}